\documentclass[journal]{IEEEtran}
\normalsize

\ifCLASSINFOpdf

\else

\fi
\usepackage[normalem]{ulem}
\usepackage{soul}
\usepackage{array}
\usepackage{color}
\usepackage{amsfonts}
\usepackage{amssymb}
\usepackage{amsmath}
\usepackage[pdftex]{graphicx}
\usepackage{cite}
\usepackage{balance}
\usepackage{caption}
\usepackage{subcaption}
\usepackage{textcomp}
\usepackage{gensymb}
\usepackage{float}
\usepackage{tabularx}
\usepackage{multirow}
\usepackage{amsthm}
\usepackage[table]{xcolor}
\usepackage{hyperref}

\definecolor{babyblackeyes}{rgb}{0.6, 0.7, 0.9}

\definecolor{aliceblack}{rgb}{0.8, 0.9, 1.0}

\usepackage[utf8]{inputenc}
\begin{document}
\title{A Vision of Self-Evolving Network Management for Future Intelligent Vertical HetNet}
\author{Tasneem Darwish, 
Gunes Karabulut Kurt, 
Halim Yanikomeroglu,  
Gamini  Senarath, 
and Peiying Zhu
}

\maketitle	

\begin{abstract}
Future integrated terrestrial-aerial-satellite networks will have to  exhibit some unprecedented characteristics for the provision of both communications and computation services, and security for a tremendous number of  devices with very broad and demanding requirements  
across multiple networks, operators, and ecosystems. Although 3GPP introduced the concept of  self-organization networks (SONs) in 4G and 5G documents to automate  network management, even this progressive concept will face several  challenges as it may not be sufficiently agile in coping with the  immense levels of complexity, heterogeneity, and mobility in the  envisioned beyond-5G integrated networks. In the presented vision, we discuss \textit{ how future integrated networks can be intelligently and  autonomously  managed to efficiently utilize resources, reduce operational costs, and achieve the targeted Quality of Experience (QoE)}. We introduce the novel concept of ``self-evolving networks (SENs)" framework, which  utilizes artificial intelligence, enabled by machine learning (ML)  algorithms, to make future integrated networks fully automated \textcolor{black}{and intelligently evolve}
with respect to the provision, adaptation, optimization, and management aspects of networking, communications, computation, and infrastructure nodes' mobility. 
To envisage the concept of SEN in future integrated networks, we use the Intelligent Vertical  Heterogeneous Network (I-VHetNet) architecture as our reference. The paper discusses five prominent scenarios 
where SEN plays the main role in providing automated network management.  Numerical results  provide  an insight on how the SEN framework improves the performance of future integrated networks. The paper  presents the leading enablers and examines the challenges associated  with the application of SEN concept in future integrated networks.


\end{abstract}
%

%
\section{Introduction}\label{Intro}

To address the ever-increasing user  demands of extremely high data rates with extremely low latency in an almost-ubiquitous manner, networks will undergo an unprecedented transformation that will make them substantially different from previous generations. This will require a radical paradigm shift in the way networks and services are managed and controlled. There is an emerging need to handle the increase in the overall complexity resulting from the transformation of networks into programmable, software-driven, service-based and holistically managed architectures, and the unprecedented agility and mobility, where users and \textcolor{black}{base stations (BSs) will be mobile in three dimensions (i.e., x, y, and z dimensions) \cite{Alzenad2019}.}



\textcolor{black}{
There is a growing employment of multi-cloud and edge computing, multiple layers of virtualization, and agile resources across different operators and ecosystems. In such environment,  obtaining a clear view of network activities is complicated and  holistically understanding the network stack is very difficult. With the introduction of such high complexity in future network environments, classical networks' control and management solutions may not be sufficient. In addition, such solutions lack network self-awareness, learning capabilities, and the ability to interact with the environment \cite{Klaine2017}. }


In this paper, we provide a vision on \textit{how future networks can be managed in order to efficiently utilise resources, reduce operational cost, and achieve the coveted QoE}. \textcolor{black}{We propose the  \textit{self-evolving networks} (SENs) framework that utilizes data-driven intelligent real-time control to automate  network management.} 
 For 4G and 5G, 3GPP proposed the concept of self-organizing network (SON) (Rel. 8-14)  and machine learning (ML) empowered SON (Rel. 15-16) \textcolor{black}{(3GPP TR 28.861, V16.0.0, Dec. 2019)}, which focused on the autonomous configuration, optimization, and healing of an existing network that has a predefined set of radio resources. 
 However, future networks will require a paradigm shift from classical SON, whereby the network  adapts its functions to specific environment states, into a self-evolving network that can maintain its performance, under highly dynamic and complex environments. SEN will drive the network management from self-organization to  continuous and automatic evolvement, \textcolor{black}{where even management policies can be self-adjusted, }to automatically react to unknown environments and triggers, requiring self-adaptive and resilient learning mechanisms. Unlike SON, SEN will be able to self-manage a network of networks that spans across multiple operators and ecosystems (e.g., satellite, aerial, and terrestrial networks). In addition, SEN will resolve conflicts and manage the coordination among the several entities in the future integrated networks. Moreover, SEN will consider the provision, optimization, and management of both communication and computational resources. Table \ref{SONSENComparison} provides a comparison between SON and SEN concepts.

\begin{table}[t]
\caption{A comparison of SON and SEN characteristics} 
\begin{tabular}{|l|l|l|}
\hline
\cellcolor{babyblackeyes}\textbf{Characteristic } &  \cellcolor{babyblackeyes}\textbf{ SON }	 & \cellcolor{babyblackeyes}\textbf{ SEN } \\ 
\hline 
\hline
\cellcolor{aliceblack}\textbf{Scope }& \parbox{3cm}{Operates in designated network or operator.} & \parbox{3cm}{Across networks, operators, and ecosystems.}\\	
\hline
\textcolor{black}{\cellcolor{aliceblack}\textbf{KPIs}} & \textcolor{black}{\parbox{3cm}{KPIs are optimized individually within a single network or operator scope.}} & \textcolor{black}{\parbox{3cm}{Many different and possibly conflicting KPIs are jointly optimized across networks, operators, and ecosystems.}} \\
\hline
\textcolor{black}{\cellcolor{aliceblack}\textbf{\parbox{1.6cm}{Capacity and coverage optimization}}} & \textcolor{black}{\parbox{3cm}{Self-establishment of 3GPP network function, resource allocation, load balancing, interference coordination,  and random access optimization within a single network or operator.}} & \textcolor{black}{\parbox{3cm}{The extra functionality of SEN includes automated dynamic network expansion and temporary network provision by controlling the mobility of infrastructure nodes, and integrated cross network optimization and resource management (i.e., selection of drones, HAPS, or satellite for data offloading).}} \\
\hline
\cellcolor{aliceblack}\textbf{\parbox{1.6cm}{Intelligence deployment}} & \textcolor{black}{\parbox{3cm}{Intelligence is applied in centralised or semi-centralised manner (within a network or ecosystem). Adds intelligence at the network edge and core.}} & \textcolor{black}{\parbox{3cm}{Fully distributed across networks, operators, and ecosystems, where edge computing and UEs resources are used as well. Integrates intelligence into the fabric of future VHetNet.}} \\
\hline
\cellcolor{aliceblack}\textbf{\parbox{1.6cm}{Coordination and conflict management}}& \parbox{3cm}{Lacks coordination and conflict avoidance among autonomic management functions of a SON. }& \parbox{3cm}{Ensures conflict-free and coordinated inter-working of multiple autonomic functions and multiple SONs that operate simultaneously in the same or interacting networks, operators, and ecosystems.}\\
\hline
\cellcolor{aliceblack}\textbf{\parbox{1.6cm}{Level of security \& privacy}}& \parbox{3cm}{Security \& privacy can be managed within a network or operator domain.} &  \parbox{3cm}{Security \& privacy will be managed across networks, operators, and ecosystems.}\\
\hline
\end{tabular}
\label{SONSENComparison} 
\end{table}

\textcolor{black}{SEN framework comprises three key components: (a) SEN engine, (b) conflict avoidance and coordination management entity, and (c) distributed and collaborative computing. }Artificial intelligence, enabled by ML algorithms, will work as the core of SEN engine and will be powered by both the communications environment's collected data (e.g., spatial and temporal traffic distributions, user preferences, and mobility patterns) and external sources improvements such as novel technologies, emerging network components, and advanced communication services. We expect that ML will be effective in learning from experience and detecting changes. Thus, with such knowledge and self-awareness, continuous, intelligent, and automated decision-making can be made to evolve the network. 
 For example, intelligent decisions can be made to inject more communications/computation resources/components, or add services in the network when there are expected demands for extra high data rates or ultra low latency edge computing. Fundamentally, SEN will support communication networks through intelligent and automated management, and communication networks will support the self-awareness, and the distributed and collaborative computing in SEN.

We utilize the \textit{intelligent vertical heterogeneous network (I-VHetNet)} architecture, which is an extension of the work in \cite{Alzenad2019}, as a reference architecture to reflect the concept of SEN on future integrated networks. 
Figure \ref{proposed_integrated_network} shows the SEN framework integration with I-VHetNet. \textcolor{black}{Unlike  SON,  whereby  the  network adapts its functions to specific environment states, \textcolor{black}{SEN conceptualizes self-evolving  network of networks  that  can  maintain}  its  performance,  under highly  dynamic  and  complex  environments.
The contributions of this paper are as follows:
\begin{enumerate}
    \item We provide a  vision on intelligently automating the service management and network operation of future integrated networks by introducing the  SEN  management framework,  which has two main components.
    \begin{enumerate}
        \item Conflict avoidance and coordination management entity, which aims to automate the process of conflict avoidance/resolution and coordination management among the distributed individual network entities \textcolor{black}{across multiple networks, operators, and ecosystems.}
        \item SEN evolution engine, which performs a continuous process of learning and monitoring of the integrated network environment, adapting to changes, and interacting with the environment to satisfy user requirements and utilize network resources efficiently.
    \end{enumerate}
    \item We discuss the integration of SEN concept with future integrated network through considering the I-VHetNet as an example architecture. 
    \item We introduce five scenarios where SEN  plays  a  vital  role  in  providing  automated  network management in future integrated networks.
    \item We present simulation results for a simple yet non trivial system to compare the performance of SON and SEN for data offloading and computing in future networks.
    \item The  main  enablers  and  challenges  associated  with  the integration of SEN in future networks are highlighted.
\end{enumerate}
}

In the next section, we introduce the concept of SENs and discuss how it differs from SONs. 
Section \ref{sec:III}, presents an overview of I-VHetNet. In Section \ref{sec:IV}, SENs enablers are discussed, and Section \ref{sec:V} presents the envisioned five scenarios, where SEN framework plays a vital role in providing automated network management along with numerical results. Several critical issues and challenges are discussed in Section \ref{challenges}, and our conclusions are presented in Section \ref{S:con}.

\section{\textcolor{black}{Self-Evolving Network Framework}}\label{SENnotSON}

Future integrated networks will provide immense heterogeneous communication and computation resources. \textcolor{black}{However, it will be almost impossible to achieve the coveted key performance indicators (KPIs) across integrated networks without intelligent and fully automated management of network services and operation. There will be many different and possibly conflicting KPIs of different operators, networks, and ecosystems.} 

\begin{figure}
\centering
\includegraphics[width=0.5\textwidth]{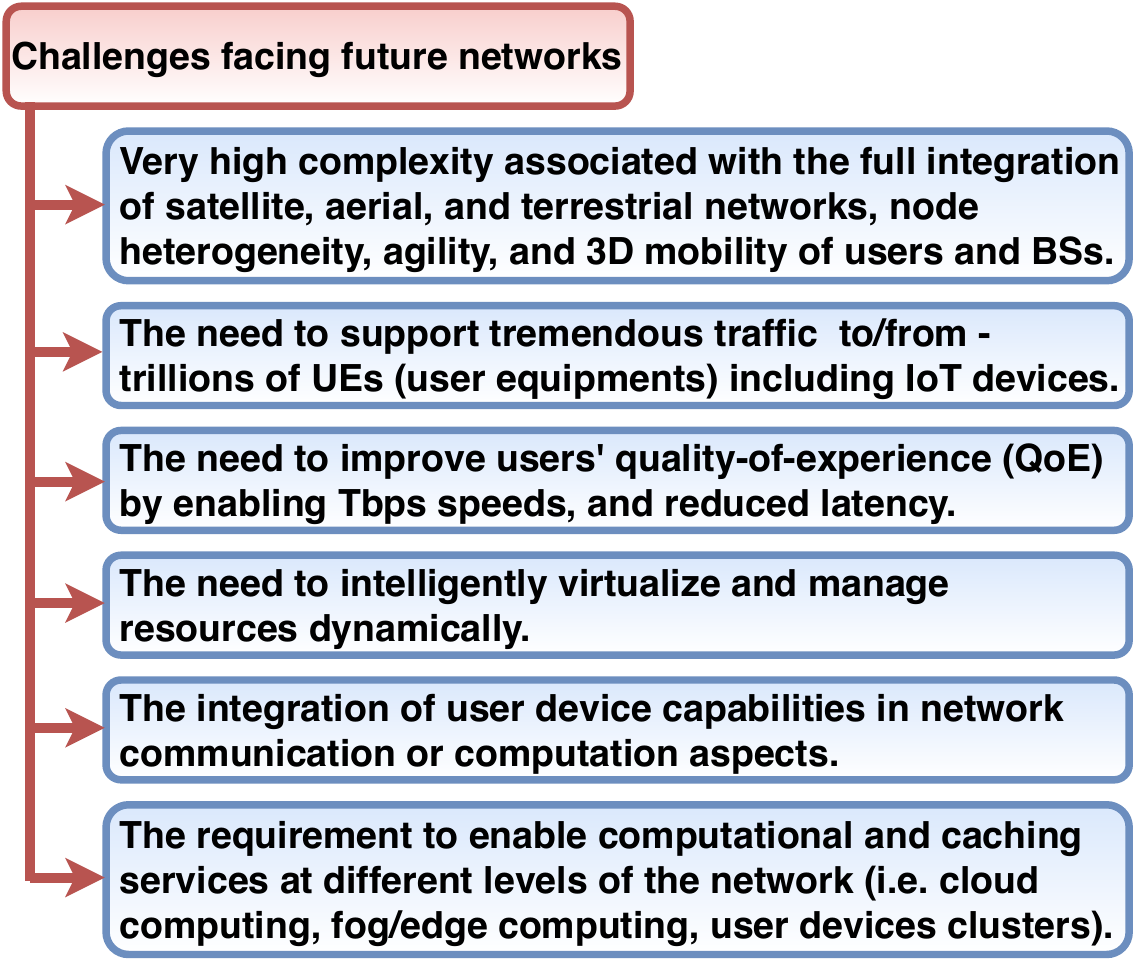}
\caption{Challenges facing future networks.}
\label{Figchallenges}
\end{figure}

Recently, some emerging ideas call for the utilization of AI/ML in managing services and network operation. The knowledge defined network concept introduced in \cite{mestres2017knowledge} utilizes Software-Defined Networking (SDN) and Network Analytics  to facilitate the adoption of AI techniques in the context of network operation and control. The ETSI Zero-touch network and Service Management (ZSM) group is formed to accelerate the definition of the required architecture and solutions in order to achieve full end-to-end automation of network and service management in the context of 5G.  3GPP SON concept 
provides the capabilities of self-configuration at the network deployment phase, self-optimization of network parameters, and self-healing to prevent/detect/correct network failures \cite{Moysen2018}. SON presents significant limitations relative to the challenges facing future networks. The challenges are summarized in Figure \ref{Figchallenges}.  Table \ref{SONSENComparison} provides a comparison between the characteristics of SON and SEN.  

To fulfill the requirements of future integrated networks, network management should go beyond the concept of executing the pre-defined management functions and should be able to automatically react to unknown environments and triggers. In SEN, network operation and service management automation will evolve through time. The network will not only learn the new environment but it will also be able to learn how to learn and what to learn. 

\begin{figure*}[tb]
\centering
\includegraphics[width=1.0\textwidth]{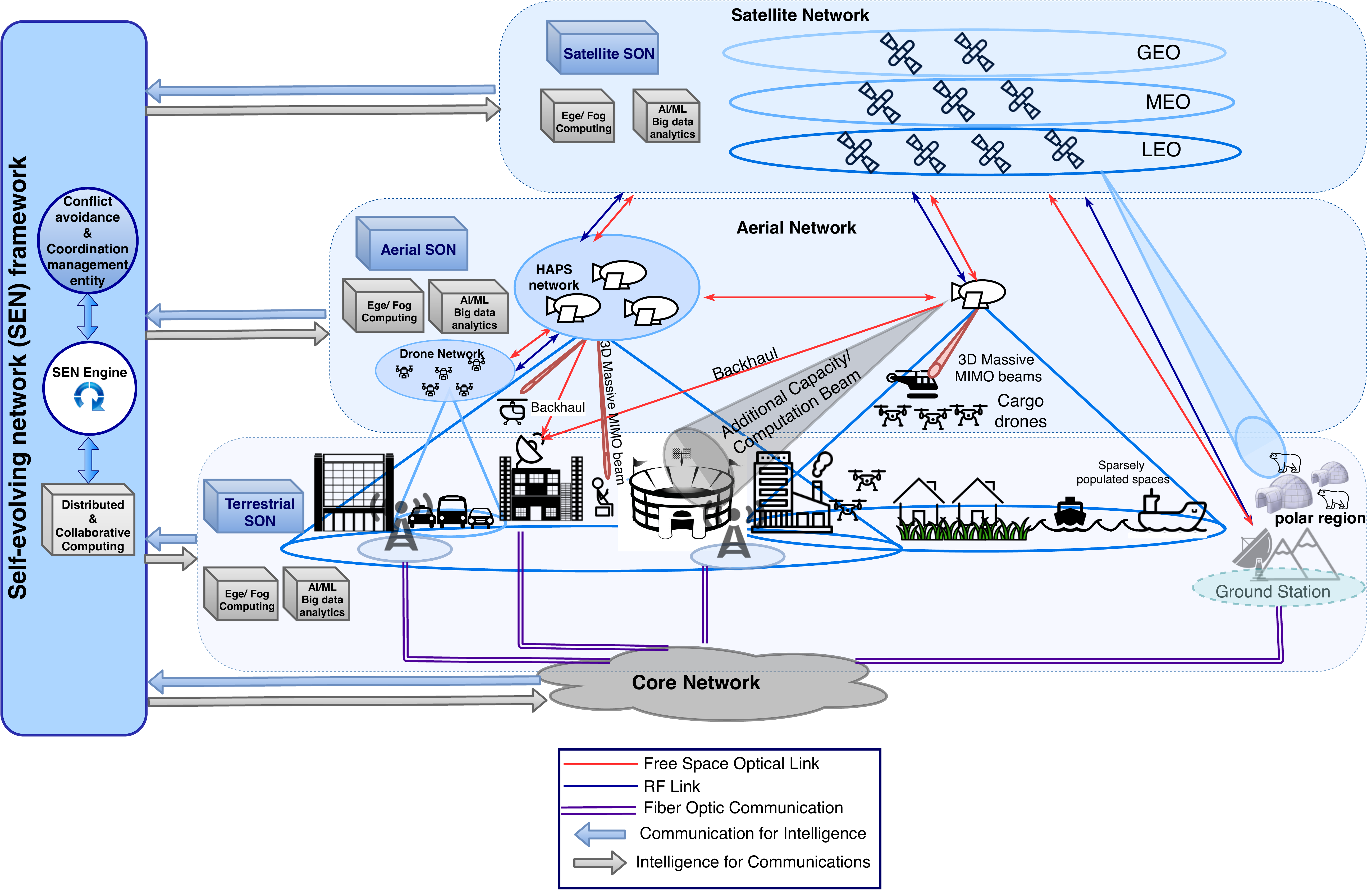}
\caption{The integrated SEN framework with the Intelligent vertical heterogeneous network (I-VHetNet) architecture. The terrestrial layer  consists of the conventional  BSs. 
UAVs, flying aircrafts, and high altitude platform station (HAPS) systems are the main components of aerial networks. UAVs can be used either as an aerial BS or as user equipment. The I-VHetNet architecture not only integrates the terrestrial-aerial-satellite networks, but it also incorporates intelligence and provides a computation and caching platform to enable multi-level edge computing. 
The distributed computing resources in I-VHetNet \textcolor{black}{form the core of the} \textcolor{black}{\it{Distribute and Collaborative Computing}}  \textcolor{black}{component of SEN framework.} 
The SEN framework \textcolor{black}{utilizes such resources to fulfil the computational requirements of the SEN evolution engine (i.e., to execute ML algorithms).} 
}
\label{proposed_integrated_network}
\end{figure*}

\textcolor{black}{SEN framework has three main components: (a) conflict avoidance and coordination management entity, (b) SEN evolution engine, and (c) distributed and collaborative computing. Figure \ref{proposed_integrated_network} shows the SEN framework and its interaction with the future integrated networks (e.g., I-VHetNet).} 
The concept of SENs implements multi-level intelligent network management policies,  which can perform across different networks, operators, and even ecosystems (e.g., cellular or satellites ecosystems). SEN concept is supported by advances in ML (e.g., federated learning, online learning, continual learning), the availability of edge and distributed collaborative computing, the agility and mobility of network resources, and the softwarization of network resource management. 

\textcolor{black}{The first component of SEN, the conflict avoidance and coordination management entity, works on multi-levels spanning from single network domain to multiple operators and ecosystems. The main role of this component is to resolve conflicts and manage the coordination among the individual network entities (microscopic level) while they are interacting with each other in a distributed peer-to-peer fashion, which will form the evolving structure and functionality of the overall SEN system (macroscopic level).} 

\begin{figure*}[tb]
\centering
\includegraphics[width=0.9\textwidth]{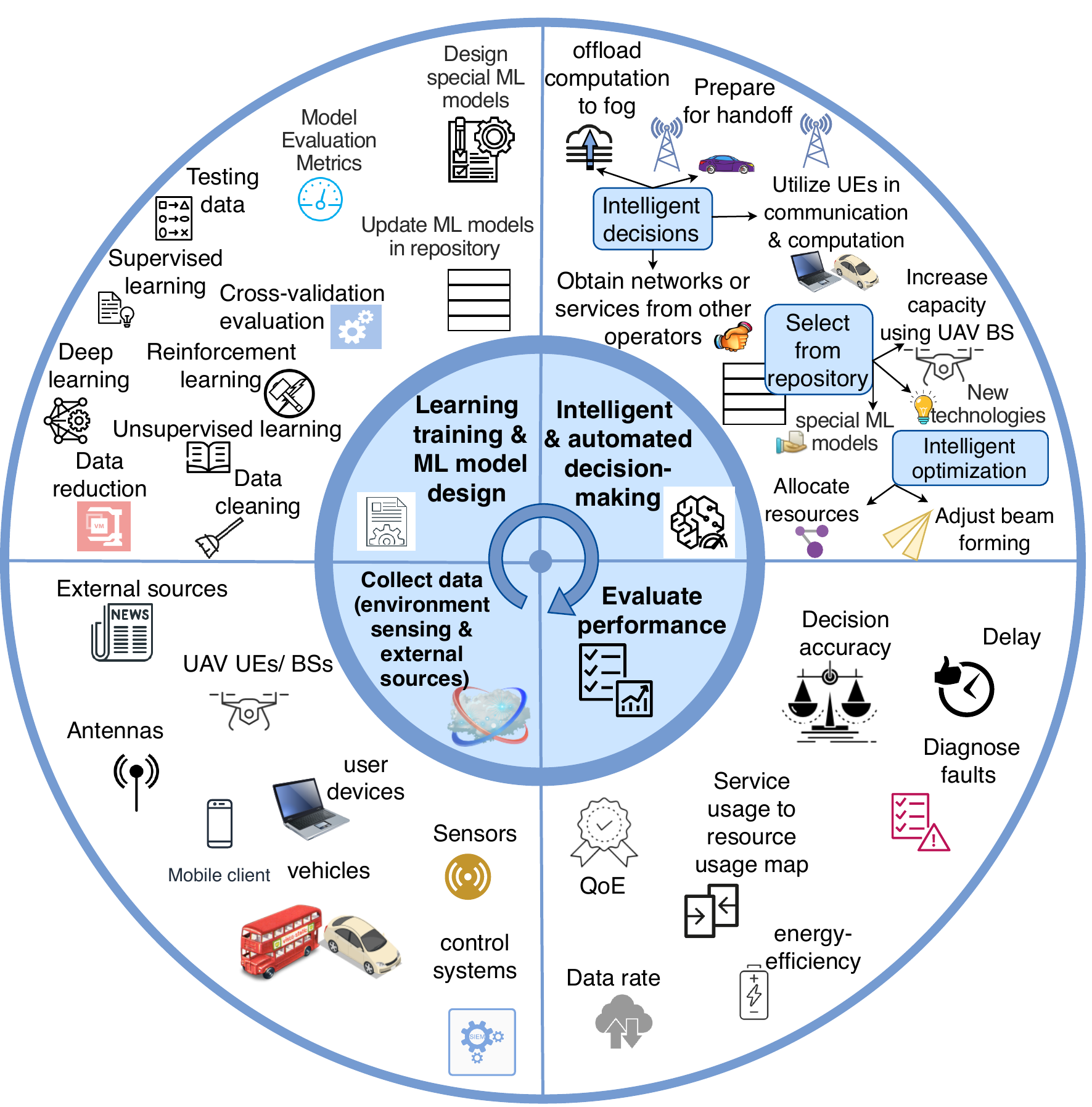}
\caption{The evolutionary cycle of SEN's evolution engine.}
\label{wheel}
\end{figure*}

\textcolor{black}{The second component of SEN is the evolution engine shown in Figure \ref{wheel}.} The engine cycle  starts by collecting data, such as network status, data traffic, and mobility patterns, through users and network devices, sensors, and external sources (e.g., news and weather forecast). The massive collected data may go through some pre-processing procedures (e.g., cleaning, reductions, transformations). 
SEN engine's core utilizes both special pre-designed ML models that can be obtained from the SEN repository and adapted models that can be modified on the spot to meet new requirements.  
In the dynamic environment of wireless networks, fast online learning algorithms executed at the network edge and distributed among UEs will be necessary to provide fast intelligent and adaptive responses to delay-sensitive applications. Offline ML is important for prediction and planning purposes. 
Afterwards, the selected/designed ML model can be used to make automated and intelligent decisions, such as automatically allocating or retrieving resources, ordering a UAV-BS, forming a new temporary network, adjusting beamforming parameters, preparing for handoff, and offloading computations to fog/cloud computing. 
Periodically, network performance and user satisfaction are measured 
then the network can intelligently decide to perform a new cycle of self-evolution.


To cope with changes in the network environment, SEN has a development repository of new adopted technologies, agile and mobile resources, coordination schemes, network components, services, intelligent decision-making models, and specially designed ML algorithms. SEN performs a continuous examination, scanning, assessing and predicting changes in application/service requirements, users' needs, and network status. When a need or a change is detected, SEN selectively obtains the suitable development tool from the repository 
and adaptively exploits it to meet the variable requirements. Basically, SEN has the characteristic of "autonomous driving networks".



\textcolor{black}{The third component of SEN framework is the distributed and collaborative computational resources which is necessary to support the computational requirements of SEN. Such computational resources are distributed vertically through the three layers of integrated networks (i.e., terrestrial, aerial, satellite), and are also distributed from the core network level to the network edge and even to the UEs level.} The distributed interaction of SEN entities eliminates the effect of single point of failure, and the system can repair or correct damages without external help. The combination of the adaptability of SENs with their distributed nature presents two major advantages: robustness against failure and scalability. The continuous evolution of SENs increases the reliability of the network. 

SENs can provide end-to-end network automation that is not limited to optimizing network configuration parameters, but can reach the level of automatically forming a temporary communications network (i.e., through mobile and agile BSs) to fulfil the  demands of a specific area for a certain time. Through ML, SENs can enable automated SDN reprogramming,  network function virtualization (NFV), and dynamic network slicing (NS) \cite{khamse2021agile}. 

\textcolor{black}{Section \ref{sec:III} explains where SEN framework can be used and how it can be integrated with future networks architecture (e.g., I-VHetNet). Section \ref{sec:IV} discusses the SEN concept enablers in future integrated networks. Section \ref{sec:V} gives examples on how SEN framework can support real scenarios.}

\section{
\textcolor{black}{SEN framework integration with I-VHetNet}}\label{sec:III}





 
\subsection{\textcolor{black}{The Interaction between SEN Framework and I-VHetNet Game-Changing Components}} 
\textcolor{black}{In this section we discuss how SEN framework supports I-VHetNet key components and how such components serve the computational and communication aspects of SEN. In this paper I-VHetNet is used as an example of future integrated network architecture. }

\paragraph{UAV as a BS} 
     Under time and energy constraints of fast UAV-BS deployment \textcolor{black}{(i.e., UxNB in 3GPP TS22.125 and TS22.261)}, several tangled complex decisions must be made very fast, including load balancing, radio resource management, route management, and beamforming. 
     \textcolor{black}{SEN intelligent and automated management approach enables the  self-deployment of UAV-BSs, handles their fast mobility and handoffs, and manages the connections of the hundreds or possibly thousands of users that are served by the UAV-BS.} 

    \paragraph{UAV as a User Equipment (UE)} Supporting thousands or millions of UAV-UEs (e.g., cargo  drones) challenges the scalability of existing solutions \cite{Bor2019}. 
   SEN will play an important role in  automating and optimizing the processes of radio resource management, and mobility management across networks and operators for such a huge number of UAVs.

    \paragraph{High Altitude  Platform Station (HAPS) Systems} 
    The HAPS has emerged as a viable quasi-stationary aerial network component 
    \cite{kurt2020vision}. 
    With free-space optical (FSO)  communications, several HAPS systems can form a  powerful backbone network and enable an ultra low latency backhaul connectivity for UAVs and various aerial network elements. Manual or semi-automated management in such a complicated system will limit its capabilities, waste its resources, and increase its operational costs. Therefore, SEN automated management is vital for HAPS systems.  \textcolor{black}{In addition, HAPS can be equipped with computational devices to support the distributed and collaborative computing component of SEN framework.} 

    \paragraph{LEO Satellites} 
   In the near future an immense number of low earth orbit (LEO) satellites are going to be orbiting the earth to provide global connectivity and Internet access to users everywhere. 
\textcolor{black}{SEN automated management services will support satellite networks requirement to be self-controlled and self-managed with automated decision-making capabilities. Likewise, satellite global coverage will support the multi-level data collection and computational offloading required for SEN functionalities. } 

\subsection{\textcolor{black}{SEN Framework Process Description in I-VHetNet  Environment}} 
 \textcolor{black}{To describe the process in which the SEN framework and I-VHetNet work together, consider the situation of providing communication coverage for a certain event in an area where there is no terrestrial network coverage. First, using the collected data about users mobility, behavior, and used applications, the SEN evolution engine uses ML to predict the location where extra network capacity and extended coverage are required. Second, using the prediction result, SEN engine makes an intelligent and automated decision to send some UAV-BSs or adjust a HAPS beam to extend the network coverage to the event area and boost its capacity to provide services for users during the event time. Through learning the QoS requirements of users applications, the SEN engine can select the best way to backhaul the UAV-BSs through terrestrial, aerial, or satellite networks. If some user applications require computational offloading, then the SEN engine directs the offloading to the most suitable computational level (e.g., cloud computing, fog computing, user equipment collaborative computing). Third, the SEN engine keeps monitoring and evaluating the network environment through collecting data about network performance and measuring user satisfaction. Fourth, the network environment evaluation is used as feedback to the SEN engine in order to adapt to changes, make more accurate intelligent and automated decisions, and evolve the network performance. Throughout this process, the conflict avoidance and coordination management entity resolves any conflicts that might arise between different components, networks, or operators. The required computational power to run the SEN engine is provided through the distributed and collaborative computing component that utilizes the available computational resources of I-VHetNet. }   

\subsection{\textcolor{black}{Distinctive Characteristics of  Integrating SEN Framework with I-VHetNet}} 
 
    
    \paragraph{A Group of Self-Evolving Networks with Distributed and Intelligent Decision-Making} \textcolor{black}{With the integration with SEN framework, I-VHetNet will consist of a group of SENs} that collectively form a large integrated SEN, which can create, organize, control, manage, and sustain itself autonomously by using the SEN framework components. This will create high adaptability to changes in the network environment and increase the  scalability, robustness, and fault-tolerance. 
    
    \paragraph{ Multi-Level Computing and Caching} 
     computational and caching capabilities are provided at multiple levels to serve future applications that require high computational capabilities. The cloud level provides the highest computational power and storage capacity. The network-edge level supports delay-sensitive applications through mobile-edge/fog computing. 
     \textcolor{black}{ The lowest level is the level of UEs (e.g., autonomous vehicles and cargo drones), where SEN framework  manages the collaboration among such smart devices and with the edge computing nodes to achieve distributed intelligent learning and decision-making.} The  computational and caching resources not only supports user applications, it also supports the intelligent automation functionality in SENs. 
     On the other hand, the SEN framework can automatically manage and self-allocate the required communications and computational resources to fulfil the constantly changing user demands. 

    \paragraph{Dynamic, 3D, and Agile Topology} in I-VHetNet 3D topology everything can move including BSs (e.g., UAV \& satellite). \textcolor{black}{ The integration with SEN framework will} 
     allow forming, splitting, and slicing of networks based on changes in user demands. 
    With the agility and flexibility of SENs, we do not need to over-engineer or excessively densify the terrestrial network to provide high throughput rates or coverage, which are necessary only for a short time (often unpredictable). \textcolor{black}{SEN evolving characteristic can intelligently and automatically manage the I-VHetNet's resources and topology to cope with variable user demands and changes in network environment.} 
    
    \paragraph{Seamless Connectivity Anywhere, Anytime, and for Everything} 
    I-VHetNet architecture extends the coverage of communications networks not only to the entire globe but also to the surrounding air and space. 
    
    
    \textcolor{black}{With SEN framework support, the mobility of resources and users can be intelligently managed to achieve seamless connectivity by realizing full coverage and required communication capacity anywhere, anytime, and for everything. On the other hand, the global connectivity of I-VHetNet supports the communication requirements of SEN framework and provides a multi-level platform for network data collection.}



\section{Self-Evolvement Enablers in future networks} \label{sec:IV}

\subsection{Massive Volume of Data} 
Massive volumes of data will be generated from sensors, surveillance cameras, smart gadgets, vehicles, UAVs, HAPS systems, and satellites. 
Such data can be used to reveal trends, hidden patterns, unseen correlations, and achieve automated decision making. It can also be used to continuously learn about 
user behaviour and enable the network to proactively adapt to changes in the communications environment. However, data anonymization techniques  are essential to maintain user privacy. 

\subsection{Softwarization Paving the Way for Intelligence}


Softwarization is expected to bring the benefits of programmability into network management and control \cite{Wang2017}. By incorporating intelligent decisions into  network softwarization, this moves the network control and management to the intelligentization dimension. For example, intelligent SDN can be reprogrammed automatically and dynamically on the basis of intelligent decisions to adapt to changes in the communications environment. 
 


\subsection{ML Science Advances}

A number of  powerful ML algorithms, such as deep neural network  
and reinforcement learning, resemble the human brain learning process of trial and error. 
In addition, new ML algorithms are emerging such as meta learning and continual learning, where a dynamic ML model can be modified and adapted through re-configuring some parameters. Moreover, research is progressing on the concepts of  learning how to learn and what to learn.
 
 For resource-limited UEs, some simplified novel ML algorithms have been proposed. FastGRNN and FastRNN are algorithms to implement recurrent neural networks (RNNs), and gated RNNs into tiny devices \cite{kusupati2018fastgrnn}.


\subsection{Edge and Fog Computing Capabilities}


The SEN framework's distributed and collaborative computing component utilizes the edge computing resources to realize the functionalities of SEN and
to execute 
ML algorithms on behalf of resource-limited devices such as smart phones or sensors \cite{darwish2018}. 
Offloading computations to the network edge has many advantages. First, data and computations can be processed locally which reduces the response delay and enables real-time data-driven applications \cite{chen2019}. Second, offloading to the edge reduces traffic and congestion towards cloud data centers. Third, edge/fog computing supports mobility-aware applications as it considers user mobility. Fourth, as data do not have to travel through many nodes in the network, user privacy and data security are more preserved.   

\subsection{Collaborative Computing and Distributed ML}

To realize the concept of distributed intelligence in SENs, the edge computing and the aggregated computational resources of UEs can be utilized to form a "collaborative edge/fog cluster".
ML tasks can be distributed among a group of collaborating fog nodes or UEs \cite{Hochul2019}. Federated learning techniques can provide a platform to achieve distributed  ML with high prediction accuracy in a privacy-preserving manner \cite{Zhao2020}.  

\section{Envisioned scenarios of SEN based on I-VHetNet architecture}\label{sec:V}

This section presents some scenarios to explain how the SEN framework can support I-VHetNet. 

\begin{figure*}
\centering
\begin{subfigure}[b]{.49\linewidth}
\includegraphics[width=\linewidth]{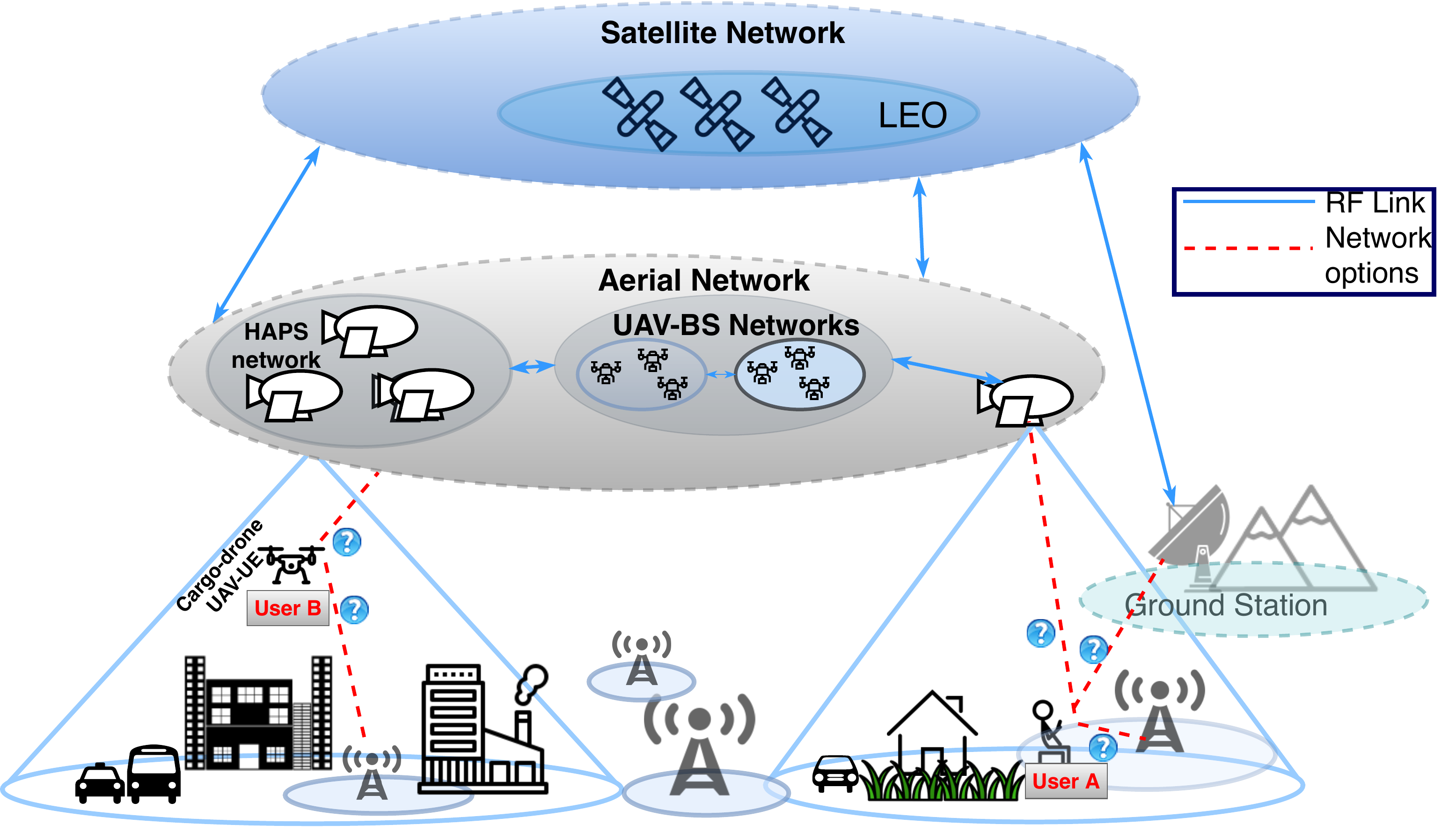}
\caption{}\label{Scenario1}
\end{subfigure}
\begin{subfigure}[b]{.49\linewidth}
\includegraphics[width=\linewidth]{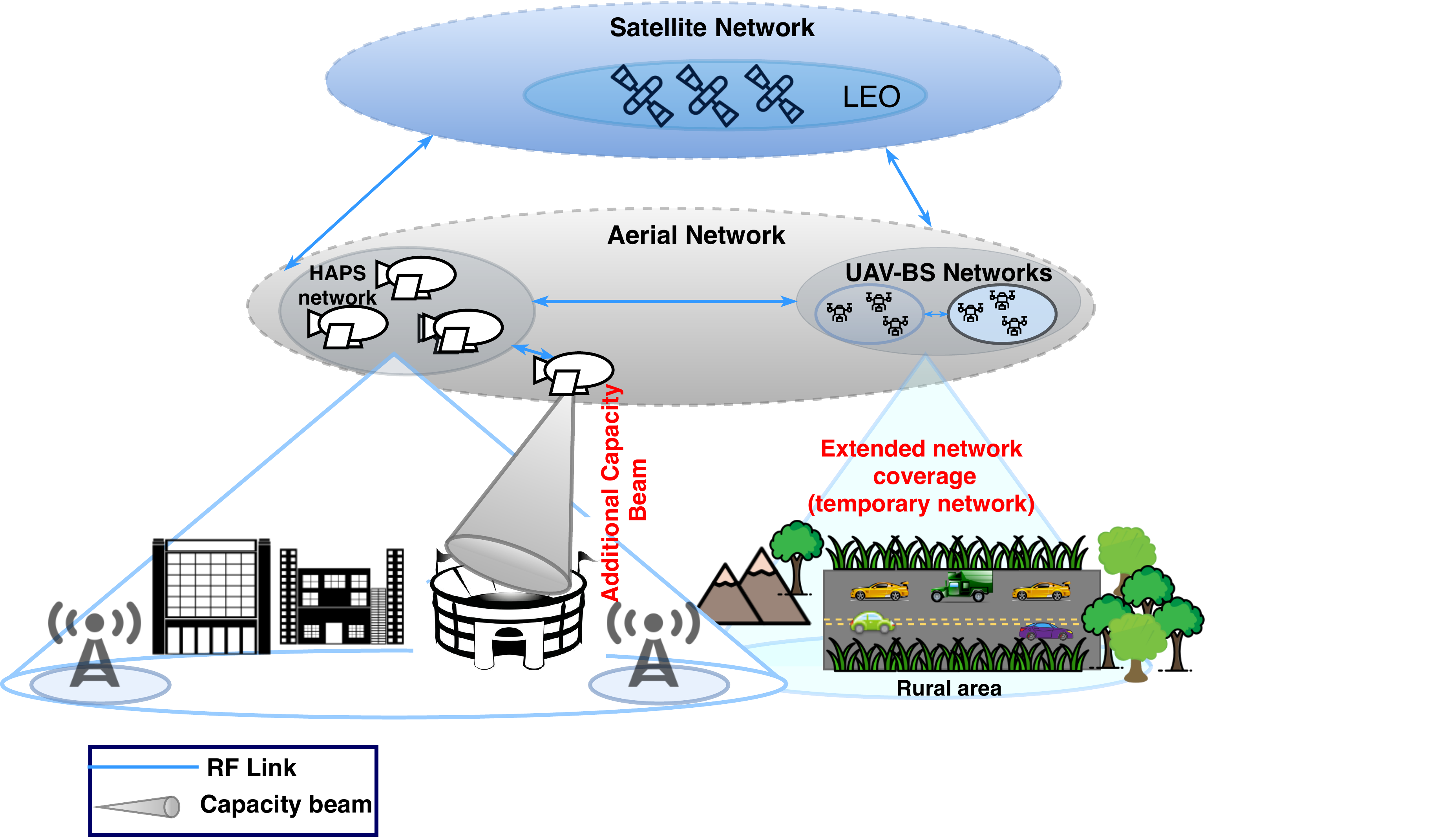}
\caption{}\label{Scenario3}
\end{subfigure}

\begin{subfigure}[b]{.49\linewidth}
\includegraphics[width=\linewidth]{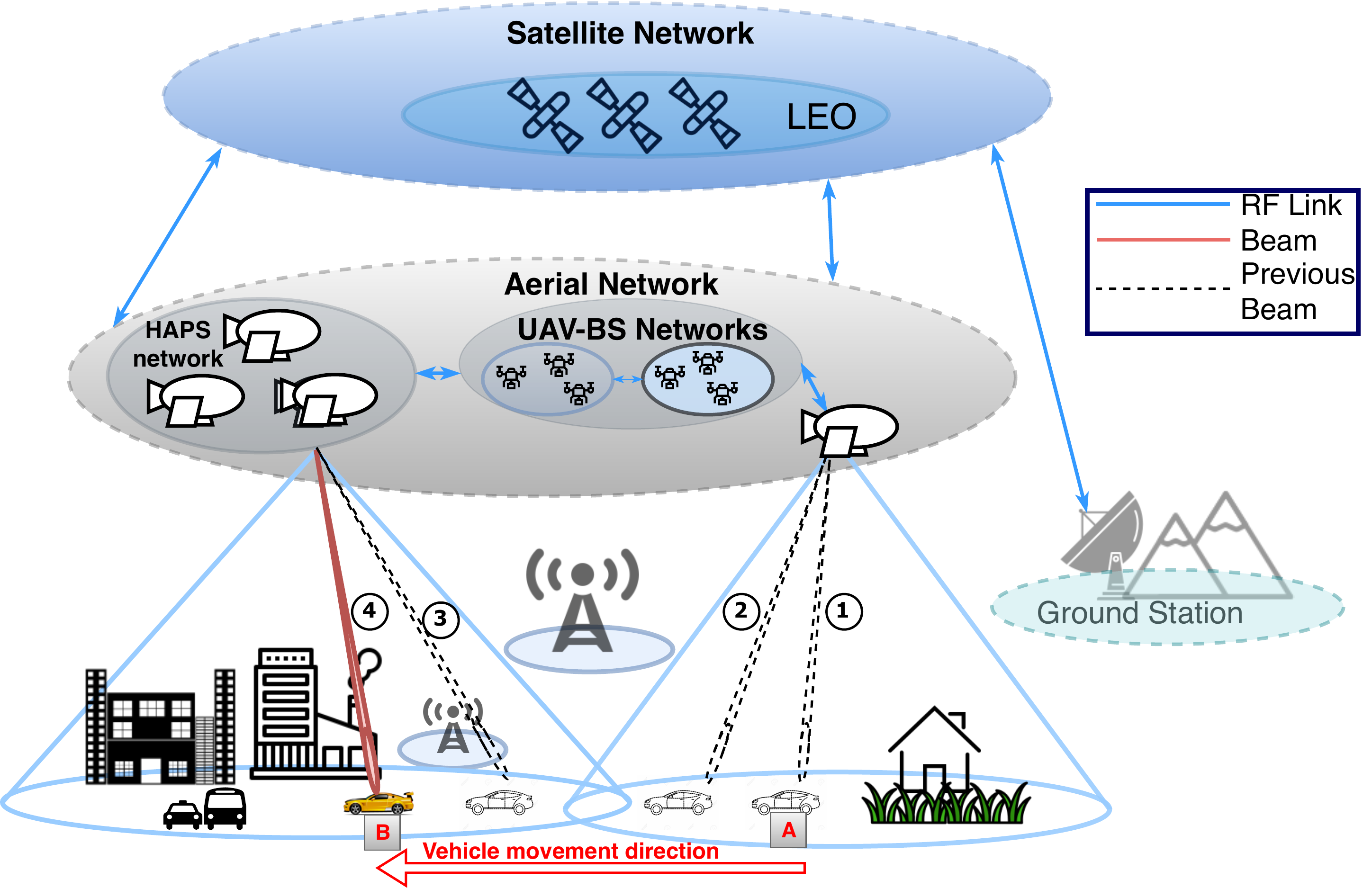}
\caption{}\label{Scenario2}
\end{subfigure}
\begin{subfigure}[b]{.49\linewidth}
\includegraphics[width=\linewidth,  height=6cm]{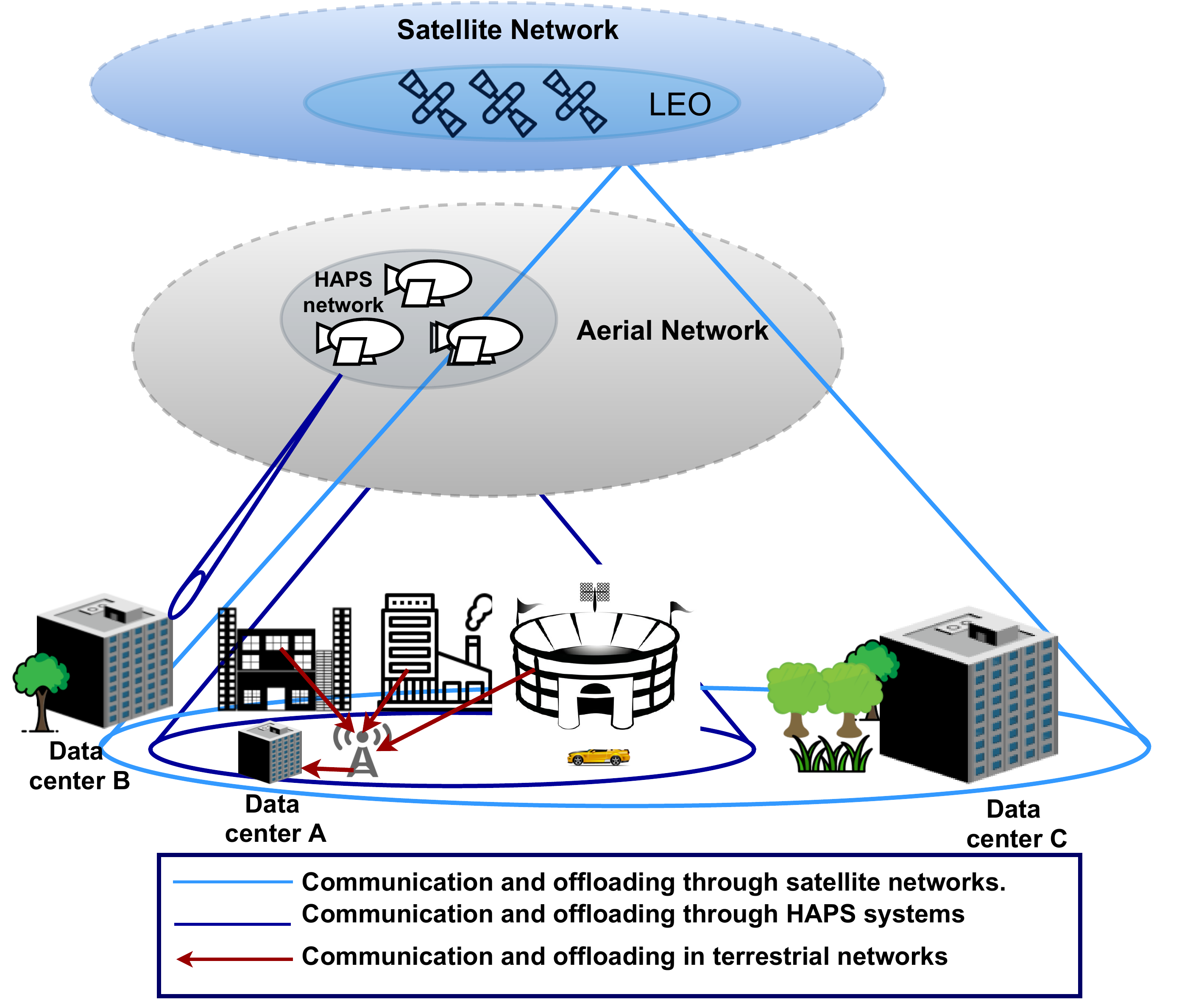}
\caption{}\label{Scenario4}
\end{subfigure}

\begin{subfigure}[b]{.49\linewidth}
\includegraphics[width=\linewidth,  height=6cm]{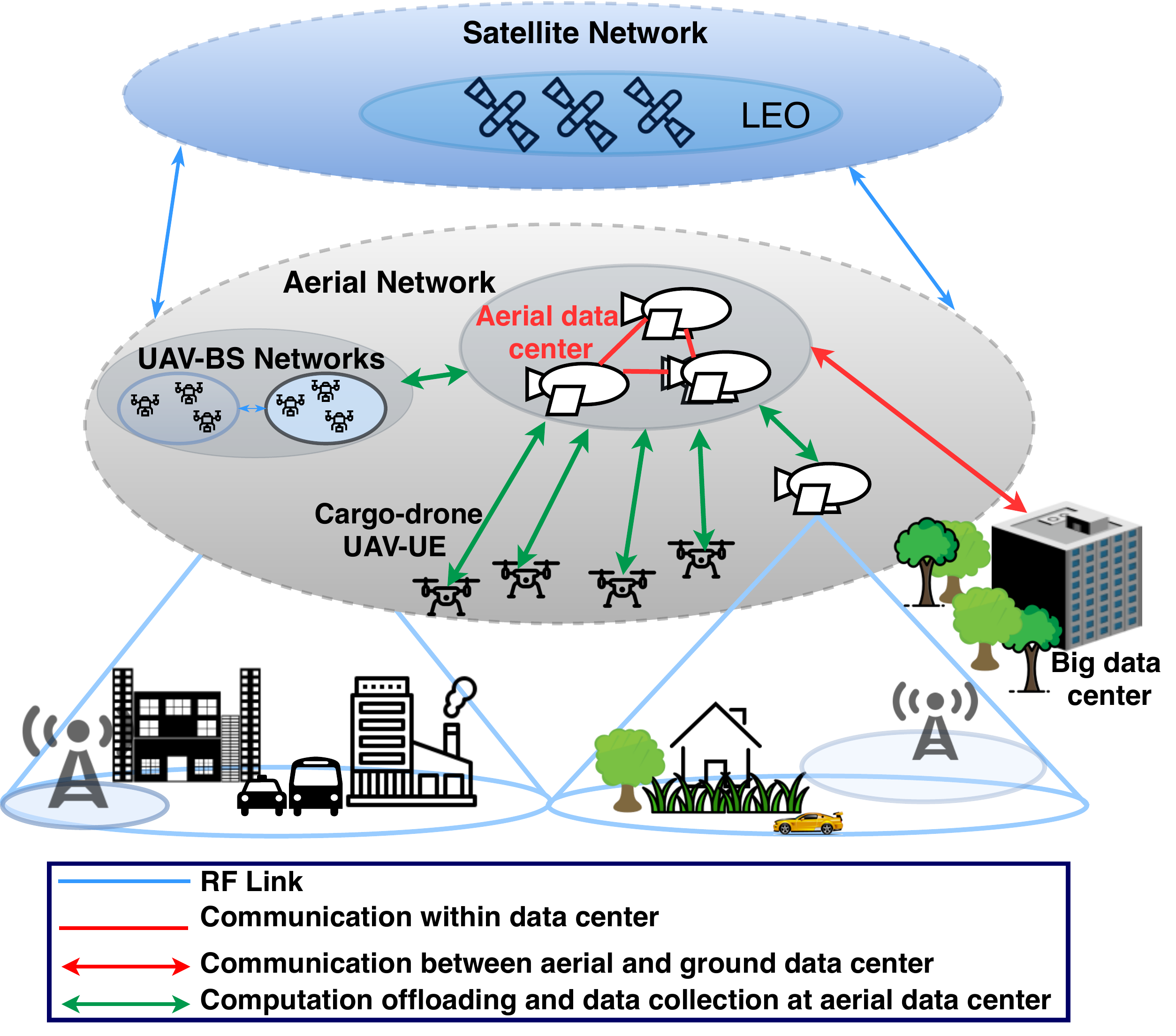}
\caption{}\label{Senario5}
\end{subfigure}
\caption{(a) \textcolor{black}{Integrated cross network optimization for users A and B. For UEs the decision on which network to connect (e.g., terrestrial, aerial, or satellite) is  complicated and requires intelligent decision-making procedure. SEN framework can autonomously  make such a decision while considering all available networks, operators, and even ecosystems.} (b) Based on intelligent prediction/detection, SEN framework can make automated decisions to extend network capacity by using a HAPS (left), extend network coverage by a temporary network of UAV-BSs (right), split a network into two or more, merge a network with other networks, and/or form a new network. Such extensions can be achieved in a timely manner and for a certain duration in order to intelligently manage network resources.  (c) Intelligent and dynamic beamforming for mobile network entities, where SEN framework can automatically adjust the formed beam through different technologies, networks, and operators to adapt to the mobility of the user. (d) Distributed data offloading and computing towards three data centers with different computing capabilities and are accessible through one of the integrated networks (terrestrial, aerial, and satellites). In this scenario, SEN framework plays the role of managing the automated and optimized distribution based on network conditions and data sizes in order to minimize the overall offloading and computing delays while utilizing network resources efficiently. (e) A HAPS network as an aerial data center, where SEN will provide the automated management of the aerial data center and coordinate data offloading and processing with the ground data center.
Due to its relatively large footprint, a HAPS network can collect data from large portions of the aerial network to use it
in supporting the self-evolution of aerial networks.}
\label{Fig:Scenarios}
\end{figure*}

\begin{figure*}
\centering
\includegraphics[width=\linewidth]{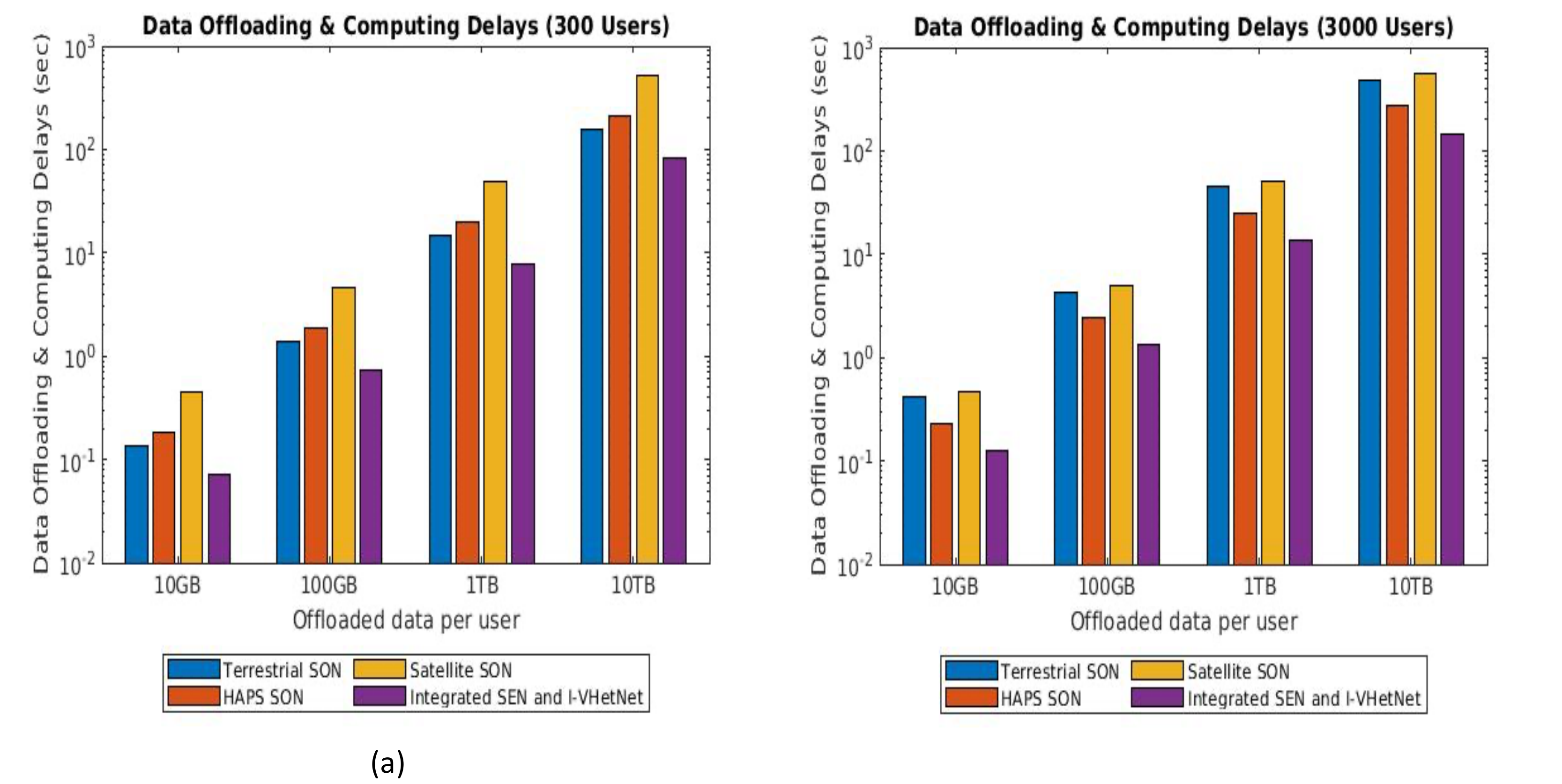}
\caption{(a) Data offloading and computing delays for 300 users based on the scenario described in Figure \ref{Scenario4}. The highest delays are when offloading the data for computing in the farthest data center C which is reachable through satellites only \textcolor{black}{(Satellite SON). In the integrated networks of I-VHetNet,  SEN framework optimally distributes the data and computation offloading among the three data centers and achieves the lowest delays (Integrated SEN and I-VHetNet).} (b) Data offloading and computing delays for 3000 users based on the scenario described in Figure \ref{Scenario4}. Compared to Figure 5 (a), it is obvious that the terrestrial data center A is overloaded, which results in extra delays (even higher than  data center B that is reachable through HAPS \textcolor{black}{(HAPS SON)). This is because SON cannot optimally utilize the resources across different networks and ecosystems. Although the number of users has increased by a factor of 10, SEN manages to achieve the lowest delays optimally distributing the offloaded data across the three layers of integrated networks.}}\label{SENResults}
\end{figure*}

\subsection{\textcolor{black}{Integrated  Cross Network Optimization}}

 Through SEN framework high load balancing across networks and BSs can be achieved by learning user demands and mobility patterns. SEN can make automated decisions of choosing the optimal serving cells, network, or even the  network components and architectures that match the required services loading situation (e.g., choosing a flat architecture, an edge/cloud server, and a suitable core network parts, etc).  
 \textcolor{black}{Using continual learning algorithms, the SEN framework utilizes the obtained  network self-awareness information to cluster} users based on their required QoE, mobility patterns, and device capabilities, and then intelligently make decisions to provide the best communications network selection. \textcolor{black}{To implement continual learning and intelligent decision-making algorithms the distributed and collaborative computing component of SEN utilizes the available edge/fog computing resources across the integrated networks. }SEN can play significant role in optimizing the handoff target and timing to guarantee seamless connection as illustrated by Figure \ref{Scenario1}. 



\subsection{Extending Network Capacity/Coverage}

I-VHetNet consists of a group of SENs. The massive collected network data can be utilized to predict future network events, whereby proactive actions can be performed to avoid delays or network failures. \textcolor{black}{For example, with deep learning (e.g., Recurrent Neural Networks (RNNs)), the spatial and temporal wireless traffic patterns can be used to match the network’s capacity to user demands by establishing a new temporary network through collaborative mobile BSs without over-engineering the network’s physical resources, as shown in Figure \ref{Scenario3}.}
\textcolor{black}{The distributed and collaborative computing component of SENs provides the required computational resources to implement such advanced learning algorithms. In addition, the conflict avoidance and coordination management entity resolves any arising conflict between the network entity while making the automated decisions.}    

\subsection{Intelligent Beamforming}

In I-VHetNet, communications will depend heavily on beamforming in order to mitigate interference \cite{vaezy2020}. 
As a SEN, multiple coordinated HAPS systems or UAV-BSs, which are  equipped with multi-antenna arrays, can form distributed MIMO-network. Through intelligent and distributed control, and with conflict avoidance entity of SEN  extremely precise beams can be created that can track the user mobility while limiting the interference, as depicted in Figure  \ref{Scenario2}. To accurately change the beam configuration based on changes in the communications network and user mobility, \textcolor{black}{reinforcement learning or continual learning approaches can help incorporate intelligence in sequential decision-making processes such as these. Such learning algorithms can be implemented and executed at the network edge under the supervision of the distributed and collaborative computing component of SEN. }

\subsection{Distributed Data Offloading and Computation through  Integrated Networks}

The SEN framework can optimally make decisions to distributed data offloading and computing across several networks, operators, and even ecosystems. For example, Figure \ref{Scenario4} describes the scenario of having three data centers, where data center A has the lowest computational capability and it is accessible through terrestrial networks, data center B has better computational capabilities but it is accessible through a HAPS network, and data center C has the highest computational capabilities  but it is accessible by satellites only. By employing the SEN framework, an optimal decision can be made to distribute the data offloading and computation across the three data centers, which are accessible through three different networks or ecosystems, taking into account the delays of both communication and computation. \textcolor{black}{To address the issue of uncertainty in the data division and distribution across the three networks, we implemented Monte Carlo simulations with 10,000 iterations. The offloading process considered erasure channels with erasure probability ranges of 0.1-0.3 (offloading through HAPS and satellite networks) and 0.3-0.5 (offloading through terrestrial networks). The maximum number of retransmission trials is set to two.  }Figure \ref{SENResults} compares the performance of SON and SEN in the data offloading and computing scenario, where SON concept is applied independently in each of terrestrial, aerial, and satellite networks while the SEN functionality can be across the integrated I-VHetNet. The simulation results show the significant SEN framework performance improvement in comparison to SON.

\subsection{Aerial Data Centers}
HAPS systems equipped with powerful processors and connected with high-speed FSO links may collectively form an aerial data center and be an aerial network intelligence enabler, as illustrated in Figure \ref{Senario5}.  \textcolor{black}{Through the distributed and collaborative computing entity of SEN framework, the  self-managed aerial data centers can provide near-user computation services for aerial networks users (e.g., drones) by allowing aerial network elements with limited resources to offload their intelligent algorithms computations.} \textcolor{black}{Through intelligent data analysis approaches, data can be analyzed in the sky to reduce response delays and decrease the burden on the aerial-to-ground communications links, which can be easily disrupted by the fast speed of aerial network elements.}  

\section{\textcolor{black}{Challenges facing the application of SEN concept in future integrated networks}}\label{challenges}
Incorporating intelligence in \textcolor{black}{future integrated networks and realizing the concept of SEN faces} many challenges that require further research work. 
\begin{itemize}
    \item \textcolor{black}{\textbf{The cost of realizing SEN concept in future integrated networks:}
    SENs depend heavily on intelligent computations performed by AI/ML algorithms. In addition, data collection across the integrated networks and distributed computing generate high communication costs.  However, using intelligent decision-making, computational resources can be placed near data sources (e.g., at the network edge) to reduce communication costs.  The appropriate selection of the required ML algorithm and its learning scope can help in reducing the computational costs. Nevertheless, further research effort is required to reduce the cost of SEN intelligent automation.}

    \item \textbf{Real-time and online learning algorithms:} Most existing ML algorithms require relatively long convergence times. In future integrated communication networks the environment may change rapidly and many applications may require fast decision-making that adapt to changes in the network. \textcolor{black}{Online and continual learning algorithms should be further enhanced to suit the scalable and highly variable network environment.}
    
    \item \textbf{ Learning what to learn and how to learn:} 
    \textcolor{black}{Learning the various states of the large scale and highly dynamic environment of future integrated networks is unfeasible. However, for a given node in a network,
    learning within certain scope and time frame is sufficient in most cases.} Choosing an appropriate data scope and duration is important to avoid learning and processing unnecessary data. This issue is quite important for network nodes with high mobility or limited processing resources. 
    To realize the concept of SEN, emerging meta learning, continual learning and the concept of learning how to learn need to be adopted and enhanced to adapt to the dynamic environment. 

   \item \textbf{Information sharing and learning across networks, operators, and ecosystems: } To implement the SEN framework, new policies are required to control data sharing and collection across different networks, operators, and ecosystems. 
   It would be a major enhancement to extend federated learning concept to work on different distributed levels and scales. 
   New schemes are needed to handle the emerging issues of data ownership rights, data credibility, data trading, and data pricing. 

 \item \textbf{Intelligent and standardized conflict resolution algorithms: } \textcolor{black}{In the environment of future integrated networks,} conflicts might arise among entities with contradicting objectives  (e.g.,  minimizing delays and sharing resources).  In  addition,  \textcolor{black}{several  operators  and  service providers are involved} where each  one  of  them  has  the goal of maximizing his own profits, which might result in greedy behavior among operators and the user might pay the price. Nevertheless, resolving conflicts among different operators, networks, or ecosystems with heterogeneous technologies and different operational policies is a very complex task. \textcolor{black}{In this domain, the deployment of SEN requires the development of intelligent conflict resolution algorithms} which can mimic the human way of thinking in similar situations. 

\end{itemize}

\section{Summary}\label{S:con} 
Providing extra resources in future integrated networks without intelligent, automated,  adaptive, and real-time control, optimization, and management may not fulfil the requirements of the emerging applications and expected QoS levels. To this  end, we introduced the self-evolving networks (SENs) framework. Empowered with AI/ML, SEN framework can make intelligent, adaptive, and automated  decisions
\begin{itemize}
\item to manage heterogeneous \textcolor{black}{network of networks}  dynamically and intelligently in  a distributed manner across different operators and ecosystems;
\item to resolve conflicts and manage coordination among several automated network entities; 
\item to satisfy the QoE requirements of an enormous number of a broad  range of UEs (including IoT devices);
\item to handle the high levels of heterogeneity, agility, and 3D mobility  of both UEs and BSs; and
\item to utilize the UE assets in  the provision of communications and computation services.
\end{itemize}
We constructed five prominent I-VHetNet communications and computation  scenarios, which model several aspects of SENs. In addition, we  listed the leading enablers and  the associated challenges.

\bibliographystyle{IEEEtran}
\bibliography{main}

\end{document}